\documentstyle[12pt]{article}
\begin{document}
\def\bea{\begin{eqnarray}}
\def\eea{\end{eqnarray}}
\title{\bf {
 The Caedy-Verlinde formula and entropy of Topological Reissner-Nordstr\"om black holes in de
Sitter spaces }}
\author{
M.R. Setare  \footnote{E-mail: rezakord@yahoo.com}
  \\
{ Department of Science, Physics group, Kordestan University,
Sanandeg,  Iran  }\\ and \\ {Institute for Theoretical Physics and
Mathematics, Tehran, Iran}\\and \\{ Department of Physics, Sharif
University of Technology, Tehran, Iran }}
%\date{\small{\today}}
\maketitle
\begin{abstract}
In this paper we discuss the question of whether the entropy of
cosmological horizon in Topological Reissner-Nordstr\"om- de
Sitter spaces can be described by the Cardy-Verlinde formula,
which is supposed to be an entropy formula of conformal field
theory in any dimension. Furthermore, we find that the entropy of
black hole horizon can also be rewritten in terms of the
Cardy-Verlinde formula for these black holes in de Sitter spaces,
if we use the definition due to Abbott and Deser for conserved
charges in asymptotically de Sitter spaces. Our result is in
favour of the dS/CFT correspondence.

 \end{abstract}
% \begin{document}
\newpage
% \vspace*{10mm}

 \section{Introduction}
Holography is believed to be one of the fundamental principles of
the true quantum theory of gravity\cite{{HOL},{RAP}}. An
explicitly calculable example of holography is the much--studied
AdS/CFT correspondence. Unfortunately, it seems that we live in a
universe with a positive cosmological constant which will look
like de Sitter space--time in the far future. Therefore, we should
try to understand quantum gravity or string theory in de Sitter
space preferably in a holographic way.. Of course, physics in de
Sitter space is interesting even without its connection to the
real world; de Sitter entropy and temperature have always been
mysterious aspects of quantum gravity\cite{GH}.\\
While string theory successfully has addressed the problem of
entropy for black holes, dS entropy remains a mystery. One reason
is that the finite entropy seems to suggest that the Hilbert space
of quantum gravity for asymptotically de Sitter space is finite
dimensional, \cite{{Banks:2000fe},{Witten:2001kn}}.
 Another, related, reason is that the horizon and entropy in
de Sitter space have an obvious observer dependence. For a black
hole in flat space (or even in AdS) we can take the point of view
of an outside observer who can assign a unique entropy to the
black hole. The problem of \ what an observer venturing inside the
black hole experiences, is much more tricky and has not been given
a satisfactory answer within string theory. While the idea of
black hole complementarity provides useful clues, \cite
{Susskind}, rigorous calculations are still limited to the
perspective of the outside observer. In de Sitter space there is
no way to escape the problem of the observer dependent entropy.
This contributes to the difficulty of de Sitter space.\\
More recently, it has been proposed that defined in a manner
analogous to the AdS/CFT correspondence,  quantum gravity in a de
Sitter (dS) space is dual to a certain
 Euclidean  CFT living on a spacelike boundary of the
dS space~\cite{Strom} (see also earlier works
\cite{Hull}-\cite{Bala}). Following the proposal, some
investigations on the dS space have been carried out
recently~\cite{Mazu}-\cite{Ogus}. According to the dS/CFT
correspondence, it might be expected that as the case of AdS black
holes~\cite{Witten2}, the thermodynamics of cosmological horizon
in asymptotically dS spaces can be identified with that of a
certain Euclidean CFT residing on a spacelike boundary of the
asymptotically dS spaces.\\
In this paper, we will show that the entropy of cosmological
horizon in the Topological Reissner-Nordstr\"om -de Sitter spaces
(TRNdS) can also be rewritten in the form of Cardy-Verlinde
formula.  We then show that if one uses the Abbott and Deser (AD)
prescription \cite{AD}, the entropy of black hole horizons in dS
spaces can also be expressed by the Cardy-Verlinde formula.
\section{Topological Reissner-Nordstr\"om-de Sitter Black Holes}
We start with an $(n+2)$-dimensional TRNdS black hole solution,
whose metric is
\begin{eqnarray}
&& ds^2 = -f(r) dt^2 +f(r)^{-1}dr^2 +r^2 \gamma_{ij}dx^{i}dx^{j}, \nonumber \\
&&~~~~~~ f(r)=k -\frac{\omega_n M}{r^{n-1}} +\frac{n \omega_n^2
Q^2}{8(n-1) r^{2n-2}}
     -\frac{r^2}{l^2},
\end{eqnarray}
where
\begin{equation}
\omega_n=\frac{16\pi G_{n+2}}{n\mbox {Vol}(\Sigma)},
\end{equation}
where $\gamma_{ij}$ denotes the line element of an $n-$dimensional
hypersurface $\Sigma$ with constant curvature $n(n-1)k$ and volume
$Vol(\Sigma)$ , $G_{n+2}$ is the $(n+2)-$dimensional Newtonian
gravity constant, $M$ is an integration constant, $Q$ is the
electric/magnetic charge of Maxwell field. When $k=1$, the metric
Eq.(1) is just the Reissner-Nordstr\"om-de Sitter solution. For
general $M$ and $Q$, the equation $f(r)=0$ may have four real
roots. Three of them are real, the largest on is the cosmological
horizon $r_{c}$, the smallest is the inner (Cauchy) horizon of
black hole, the middle one is the outer horizon $r_{+}$ of the
black hole. And the fourth is negative and has no physical
meaning. The case $M=Q=0$ reduces
to the de Sitter space with a cosmological horizon $r_{c}=l$.\\
When $k=0$ or $k<0$, there is only one positive real root of
$f(r)$, and this locates the position of cosmological horizon
$r_{c}$.\\
In the case of $k=0$, $\gamma_{ij}dx^{i}dx^{j}$ is an
$n-$dimensional Ricci flat hypersurface, when $M=Q=0$ the solution
Eq.(1) goes to pure de Sitter space
\begin{equation}
ds^{2}=\frac{r^{2}}{l^{2}}dt^{2}-\frac{l^{2}}{r^{2}}dr^{2}+r^{2}dx_{n}^{2}
\end{equation}
in which $r$ becomes a timelike coordinate.\\
When $Q=0$, and $M\rightarrow -M$ the metric Eq.(1)is the TdS
(Topological de Sitter) solution \cite{{cai93},{med}}, which have
a cosmological horizon and a naked singularity, for this type of
solution, the Cardy-Verlinde formula also work well.
\\
In the BBM prescription \cite{BBM}, the gravitational mass,
subtracted the anomalous Casimir energy, of the TRNdS solution is
\begin{equation}
\label{3eq3} E=-M =-\frac{r_c^{n-1}}{\omega_n} \left (k
-\frac{r_c^2}{l^2} +
    \frac{n\omega_n^2 Q^2}{8(n-1)r_c^{2n-2}}\right).
\end{equation}
 Some thermodynamic
quantities associated with the cosmological horizon are
\begin{eqnarray}
 && T= \frac{1}{4\pi r_c} \left(-(n-1)k +(n+1)\frac{r_c^2}{l^2}
    +\frac{n\omega_n^2 Q^2}{8 r_c^{2n-2}}\right), \nonumber \\
&& S =\frac{r_c^n\mbox{Vol}(\sigma)}{4G}, \nonumber \\
&& \phi =-\frac{n}{4(n-1)}\frac{\omega_n Q}{r_c^{n-1}},
\end{eqnarray}
where $\phi$ is the chemical potential conjugate to the charge
$Q$. \\
 The Casimir energy $E_c$,
defined as $E_c =(n+1) E-nTS-n\phi Q$ in this case, is found to be
\begin{equation}
E_c=-\frac{2nkr_c^{n-1}\mbox{Vol}(\sigma)}{16\pi G},
\end{equation}
when $k=0$, the Casimir energy vanishes, as the case of
asymptotically AdS space. When $k=\pm 1$, we see from Eq.(6) that
the sign of energy is just contrast to the case of TRNAdS space
\cite{youm}.\\
 Thus we can see that the entropy
 Eq.(5)of the cosmological horizon can be rewritten as
 \begin{equation}
 S=\frac{2\pi l}{n}\sqrt{|\frac{E_{c}}{k}|(2(E-E_q)-E_c)},
\end{equation}
where
\begin{equation}
E_q = \frac{1}{2}\phi Q =-\frac{n}{8(n-1)}\frac{\omega_n
Q^2}{r_c^{n-1}}.
\end{equation}
We note that  the entropy expression (7) has a similar form as the
case of TRNAdS black holes \cite{youm}.\\
 For the black hole
horizon, which there is only for the case $k=1$ associated
thermodynamic quantities are
\begin{eqnarray}
\label{3eq8} && \tilde T=\frac{1}{4\pi r_+}\left( (n-1)
-(n+1)\frac{r_+^2}{l^2} -\frac{n\omega_n^2 Q^2}
   {8r_+^{2n-2}}\right), \nonumber \\
&& \tilde S=\frac{r_+^n \mbox{Vol}(\sigma)}{4G}, \nonumber \\
&& \tilde \phi =\frac{n}{4(n-1)}\frac{\omega_n Q}{r_+^{n-1}}.
\end{eqnarray}
The AD mass of TRNdS solution can be expressed in terms of black
hole horizon radius $r_+$ and charge $Q$,
\begin{equation}
 \tilde E =M =\frac{r_+^{n-1}}{\omega_n} \left
(1-\frac{r_+^2}{l^2} +
   \frac{n\omega_n^2 Q^2}{8(n-1)r_+^{2n-2}}\right).
\end{equation}
In this case, the Casimir energy, defined as
 $\tilde E_c
=(n+1)\tilde E -n\tilde T\tilde
 S-n\tilde \phi Q$, is
\begin{equation}
 \tilde E_c =\frac{2n r_+^{n-1}\mbox{Vol}(\sigma)}{16\pi
G},
\end{equation}
and the black hole entropy $\tilde S$ can be rewritten as
\begin{equation}
\label{3eq11} \tilde S =\frac{2\pi l}{n}\sqrt{\tilde E_c |2(\tilde
E-\tilde E_q)-\tilde E_c|},
\end{equation}
where
\begin{equation}
\tilde E_q =\frac{1}{2}\tilde \phi Q=\frac{n\omega_n
Q^2}{8(n-1)r_+^{n-1}},
\end{equation}
which is the energy of electromagnetic field outside the black
hole horizon. Thus we demonstrate that the black hole horizon
entropy of TRNdS solution can be expressed in a form as the
Cardy-Verlinde formula. However, if one uses the BBM mass Eq.(4)
the black hole horizon entropy $\tilde{S}$ cannot be expressed by
a form like the Cardy-Verlinde formula. Our result is in favour of
the dS/CFT correspondence.
\section{Conclusion}
The Cardy-Verlinde formula recently proposed by  Verlinde
\cite{Verl}, relates the entropy of a  certain CFT to its total
energy and Casimir energy in arbitrary dimensions. In the spirit
of dS/CFT correspondence, this formula has been shown to hold
exactly for the cases of dS Schwarzschild, ds topological, ds
Reissner-Nordstr\"om and dS Kerr black holes. In this paper we
have further checked the Cardy-Verlinde formula with topological
Reissner-Nordstr\"om de Sitter black hole.\\
 For spacetimes of
black holes in dS spaces, the total entropy is the sum of black
hole horizon entropy and cosmological horizon entropy. If one uses
the BBM mass of the asymptotically dS spaces, the black hole
horizon entropy cannot be expressed by a form like the
Cardy-Verlinde formula\cite{cai93}.  In this paper, we have found
that if one uses the AD prescription to calculate conserved
charges of asymptotically dS spaces, the TRNdS black hole horizon
entropy can also be rewritten in a form of Cardy-Verlinde formula,
which indicates that the thermodynamics of black hole horizon in
dS spaces can be also described by a certain CFT. Our result is
also reminiscent of the Carlip's claim~\cite{Carlip} that for
black holes in any dimension the Bekenstein-Hawking entropy can be
reproduced using the Cardy formula~\cite{Cardy}.

\end{document}